\documentclass[letterpaper, 10 pt, conference]{ieeeconf} 

\let\labelindent\relax

\usepackage[hidelinks]{hyperref}

\usepackage{resizegather}

\usepackage{cite}
\usepackage{amsmath,amssymb,amsfonts,mathrsfs}
\usepackage{graphicx}
\usepackage{mdframed}
\usepackage{textcomp}
\usepackage{textcomp}
\usepackage{tablefootnote}
\usepackage{threeparttable}
\usepackage{bbm}
\usepackage{dsfont}
\usepackage{tikz}
\usepackage{graphicx}
\usepackage{tkz-euclide}
\usepackage{algorithm}
\usepackage{algpseudocode}
\usepackage{booktabs}
\usetikzlibrary{positioning}
\usetikzlibrary{shapes}
\usetikzlibrary{shapes.misc}
\usetikzlibrary{shapes.geometric}
\usetikzlibrary{plotmarks}
\usetikzlibrary{intersections}
\usetikzlibrary{calc}
\usetikzlibrary{fit}
\usetikzlibrary{patterns,tikzmark}
\usetikzlibrary{matrix,decorations.pathreplacing,calc}
\usetikzlibrary{backgrounds}
\usepackage{amsthm}

\tikzset{cross/.style={cross out, draw, 
         minimum size=2*(#1-\pgflinewidth), 
         inner sep=0pt, outer sep=0pt}}

 \usepackage{circuitikz} 
 
\usepackage{enumitem}

\newcommand{\ci}[0]{u}

\newcommand{\x}[0]{x}
\newcommand{\z}[0]{\mathbf{z}}

\newcommand{\vertiii}[1]{{\left\vert\kern-0.25ex\left\vert\kern-0.25ex\left\vert #1 
    \right\vert\kern-0.25ex\right\vert\kern-0.25ex\right\vert}}

\newtheorem{theorem}{Theorem}
\newtheorem{proposition}{Proposition}

\newtheorem{corollary}{Corollary}
\newtheorem{lemma}{Lemma}

\newtheorem{problem}{Problem}
\newtheorem{assumption}{Assumption}

\newtheorem{definition}{Definition}

\newmdenv[
    linecolor=white, backgroundcolor=lightgray!15, innertopmargin=5pt, innerbottommargin=5pt, skipabove=10pt, skipbelow=10pt
]{graybox}

\makeatletter
\renewcommand{\fps@figure}{htp}
\renewcommand{\fps@table}{htp}
\makeatother

\newcommand{\R}{\mathbb{R}}

\usepackage{comment}
\definecolor{blue(munsell)}{rgb}{0.0, 0.5, 0.69}

\newcommand{\parstart}[1]{\noindent \textbf{#1.}\;}

\usepackage[stretch=10,shrink=20,step=5]{microtype}
\usepackage{titlesec} 
\titlespacing*{\section}{0pt}{*0.4}{*0.4}
\titlespacing*{\subsection}{0pt}{*0.3}{*0.3}
\makeatletter
\def\thm@space@setup{%
  \thm@preskip=2pt plus 1pt minus 1pt%
  \thm@postskip=2pt plus 1pt minus 1pt%
}
\makeatother

\AtBeginDocument{%
  \setlength{\textfloatsep}{1pt plus 1pt minus 2pt}
  \setlength{\floatsep}{2pt plus 1pt minus 1pt}
  \setlength{\intextsep}{2pt plus 1pt minus 1pt}
  \setlength{\abovecaptionskip}{2pt}
  \setlength{\belowcaptionskip}{2pt}
}

\IEEEoverridecommandlockouts
\usepackage{xcolor} 
\usepackage[normalem]{ulem}
\makeatletter

\def\@IEEEstatictitlevskip{{\normalfont\normalsize\vskip 1\baselineskip plus 0.375\baselineskip minus 0.1875\baselineskip}}
\makeatother

\title{\LARGE \bf Control Barrier Function Synthesis for Differential-Algebraic Systems}

\author{Hongchao Zhang$^{1}$, Mohamad H. Kazma$^{2}$, Ben Wooding$^{1}$, Meiyi Ma$^{1}$, Taylor T. Johnson$^{1}$ and Ahmad F. Taha$^{2}$
\thanks{$^{1}$ Department of Computer Science, $^{2}$ Department of Civil and Environmental
Engineering, Vanderbilt University, Nashville, TN 37235 USA.
{\tt\scriptsize hongchao.zhang@vanderbilt.edu}
}
\thanks{This work is supported by the NSF under Grants 2220401, 2443803, 2152450 2550934, and the DARPA under Grant FA8750-23-C-0518. Any opinions, findings, and conclusions or recommendations expressed in this paper are those of the authors and do not necessarily reflect the views of DARPA or NSF.}
}

\makeatletter
\def\ps@headings{%
  \def\@oddhead{\raisebox{0.25in}[0pt][0pt]{\hbox to\textwidth{\scriptsize\leftmark\hfil}}}%
  \let\@evenhead\@oddhead \def\@oddfoot{}\def\@evenfoot{}}
\makeatother

\begin{document}
\newdimen\origiwspc%
\newdimen\origiwstr%
\origiwspc=\fontdimen2\font
\origiwstr=\fontdimen3\font

\maketitle
\thispagestyle{headings} 
\pagestyle{empty} 

\begin{abstract}
	Control barrier certificates (CBCs) have been extensively studied for ordinary differential equation (ODE) systems, but their  synthesis to differential-algebraic equation (DAE) systems has remained an open problem. This paper presents the first  CBC synthesis framework for DAE systems, addressing safety for both index-1 and higher-index systems. We transform DAE-CBF synthesis into ODE-CBF synthesis with additional compatibility constraints and develop an alternating synthesis framework. We propose an anchored sum-of-squares (SOS) formulation that guarantees geometric correctness on the constraint manifold by construction. This formulation reduces verification to feasibility certification and enables lifting back to the full DAE state space. The approach is validated on a 24-dimensional three-link manipulator, demonstrating safe operation under algebraic constraints.
\end{abstract}


\section{Introduction}\label{sec:intro}

Safety-critical control of cyber-physical systems requires formal guarantees that the system state remains within a prescribed safe region during operation. Control barrier certificates (CBCs) ~\cite{prajna2004safety} provide such guarantees by \textit{(i)} synthesizing a Control Barrier Function (CBF) $b(\x)$, a scalar function whose superlevel set encodes the safe region, and \textit{(ii)} pairing it with a certificate triple $(b, \Xi, \gamma)$, where $\Xi$ are Positivstellensatz multipliers and $\gamma > 0$ a certificate margin, to certify forward invariance through an inequality constraint on the control input~\cite{ames2019control}. The resulting safety filter is computationally efficient, composable with nominal objectives, and extensible to higher relative degrees~\cite{xiao2021high}, making CBCs one of the most widely adopted certificate tools for real-time safety enforcement.


Existing CBF synthesis methods, however, assume ordinary differential equation (ODE) models where dynamics evolve over unconstrained differential state space. Many engineering systems are more naturally described by differential-algebraic equations (DAEs), in which differential dynamics are coupled with algebraic constraints that encode physical laws such as kinematic closure in robotic manipulators~\cite{Park2024b} and power balance in electrical networks~\cite{kundur2007power,Grob2016}. These constraints restrict the state to a manifold and introduce additional structural requirements, such as differentiation index and compatibility conditions, which are absent in ODE models.

Extending CBCs to DAE systems introduces a key technical challenge; the control input must simultaneously satisfy the CBF inequality \emph{and} preserve the algebraic constraint manifold, a \emph{compatibility} requirement absent in ODE formulations. CBFs for DAE systems were first introduced in~\cite{Park2024b}, however without accounting for compatibility, which may drive the state off the manifold and produce certificates that are not physically meaningful. In~\cite{zhang2026verification}, DAE-aware CBFs are introduced to resolve this challenge through projected dynamics on the constraint manifold, providing necessary and sufficient verification conditions for a given candidate barrier. However, the synthesis problem remains open; that is, given a DAE system and a safe set, it is not clear how to \emph{find} a valid DAE-aware CBF satisfying verification conditions.

\parstart{Prior work}
Existing barrier-certificate work for DAE systems includes safety verification~\cite{pedersen2016verification} and distributed construction of safety-preserving setpoint policies for power networks~\cite{kundu2019distributed}.
In contrast, CBF-based safe control has been studied primarily for ODE systems, including hybrid systems~\cite{ames2019control} and higher-relative-degree systems~\cite{xiao2021high}.
Synthesis of CBFs has been addressed via sum-of-squares (SOS) programming~\cite{jarvis2003control,clark2024semialgebraic}, and neural networks with verification-in-the-loop (counterexample-guided)~\cite{Edwards2024b,zhang2024seev}. However, these methods do not account for algebraic constraints that arise in DAE systems. Safety of DAEs has received comparatively little attention. Reference~\cite{Park2024b} extends CBFs to DAE systems for a flexible-link manipulator; however, no verification conditions are provided. Furthermore, reference~\cite{zhang2026verification} enforces compatibility and provides verification of DAE-aware CBFs, but left the synthesis of such CBFs unexplored.


\parstart{Contributions}
We present the first framework for synthesizing CBCs for DAE systems. The contributions are as follows.
\begin{itemize}[leftmargin=*]
    \item We propose an anchored SOS formulation (Theorem~\ref{th:cbcdae}) that structurally guarantees correctness on the constraint manifold, while Positivstellensatz certificates ensure feasibility of a compatible control input.
    \item We develop an alternating SOS program (Algorithm~\ref{alg:alternating_sos_cbf}) that jointly synthesizes a polynomial CBF and its certificates for index-1 and index-$\nu$ DAEs, initialized by the certified warm start in Proposition~\ref{prop:init}.
    \item We prove that a barrier synthesized on a reduced ODE lifts to the full DAE state space while preserving both correctness and feasibility (Corollary~\ref{thm:lifting}), and show that each semidefinite program (SDP) can be replaced by a linear program using diagonally dominant sum-of-squares (DSOS) certificates, together mitigating the scalability bottlenecks.
    \item {\spaceskip=.9\fontdimen2\font plus \fontdimen3\font minus \fontdimen4\font We validate the framework on a 24-dimensional three-link manipulator DAE demonstrating the proposed DAE-aware CBC ensures safety, whereas an ODE-only baseline violates.\par}
\end{itemize}
\vspace{-1pt}

\section{Problem Formulation}\label{sec:problem}
\subsection{DAE System Model}\label{subsec:model}
We consider a semi-explicit nonlinear control-affine DAE system with dynamic states $\x_{d}(t) \in \mathbb{R}^{n_{d}}$, algebraic states $\x_{a}(t) \in \mathbb{R}^{n_{a}}$, overall state $\x(t) = [\x_{d}^{T}, \x_{a}^{T}]^{T} \in \mathbb{R}^{n_{x}}$ ($n_{x} = n_{d} + n_{a}$), and control input $\ci(t) \in \mathbb{R}^{n_{u}}$:
\begin{subequations}
\label{eq:dyn}
\begin{align}
    \dot{\x}_{d}(t) &= f(\x) + g(\x) \ci(t), \label{eq:dyn_diff} \\
    0 &= \phi(\x), \label{eq:dyn_alg}
\end{align}
\end{subequations}
{\spaceskip=.9\fontdimen2\font plus \fontdimen3\font minus \fontdimen4\font where $f: \mathbb{R}^{n_x} \rightarrow \mathbb{R}^{n_d}$, $g: \mathbb{R}^{n_x} \rightarrow \mathbb{R}^{n_d \times n_u}$, and $\phi: \mathbb{R}^{n_x} \rightarrow \mathbb{R}^{n_m}$ with components $\phi_{\ell}$ ($\ell = 1,\ldots,n_m$) are smooth. We write $f_{d} := f$ and $g_{d} := g$ when emphasizing the differential subsystem. The constraint manifold is $\mathcal{M} := \{\x \in \mathbb{R}^{n_x} : \phi(\x) = 0\}$. The \emph{differentiation index}~\cite{Volker2005} $\nu$ is the minimum number of differentiations of $\phi$ required to obtain an ODE. We denote the Jacobians $J_{d}(\x) := \nabla_{\x_{d}} \phi(\x)$ and $J_{a}(\x) := \nabla_{\x_{a}} \phi(\x)$, and the admissible input set $\mathcal{U} := \{\ci : A_{u} \ci \le r_{u}\}$ with $n_{u}^{c}$ constraints.\par}

\subsection{Safety of DAE Systems}\label{subsec:safety}
The safe set is $\mathcal{C} := \{\x : h(\x) \ge 0\}$ for a smooth $h : \mathbb{R}^{n_x} \to \mathbb{R}$. A barrier function $b : \mathbb{R}^{n_x} \to \mathbb{R}$ defines $\mathcal{D} := \{\x : b(\x) \ge 0\}$. Safety requires positive invariance of $\mathcal{D} \cap \mathcal{M}$, i.e., $\x(0) \in \mathcal{D} \cap \mathcal{M}$ implies $\x(t) \in \mathcal{D} \cap \mathcal{M}$ for all $t \ge 0$.

For standard ODE systems, a CBF guarantees invariance of $\mathcal{D}$ via the condition $\frac{\partial b}{\partial \x}(f(\x) + g(\x)\ci) \ge -\alpha(b(\x))$, where $\alpha(\x)$ is a class-$\mathcal{K}_{\infty}$ function~\cite{ames2019control}. For DAE systems, the control input must additionally preserve the constraint manifold, leading to the projected dynamics framework developed in~\cite{zhang2026verification}.

Differentiating $\phi$ up to $k$ times along the dynamics yields the $k$-th level constraint $d^k\phi/dt^k = 0$. We denote the corresponding Jacobians $J_{d}^{(k)}(\x) := \nabla_{\x_{d}} (d^{k}\phi/dt^{k})$ and $J_{a}^{(k)}(\x) := \nabla_{\x_{a}} (d^{k}\phi/dt^{k})$, and write $(\cdot)^{\dagger}$ for the Moore-Penrose pseudoinverse. The \emph{control influence matrix} at level~$k$ is $\eta^{(k)}(\x) := (J_d^{(k)}(\x) f_d(\x))^{d'-1} J_d^{(k)}(\x)$, where $d'$ is the relative degree of the differential subsystem. The product $\eta^{(k)} g_d$ is the matrix through which $\ci$ enters the $k$-th time derivative of $\phi$; it determines whether the control can enforce compatibility at level~$k$. When the differential states are directly actuated ($d' = 1$), $\eta^{(k)} = J_d^{(k)}$.

\begin{assumption}\label{assumption:regularity}
The DAE system~\eqref{eq:dyn} satisfies: (i)~$\phi(\x)$ does not depend on~$\ci$; (ii)~the extended Jacobian $\big[J_{a}^{(\nu)}(\x), \eta^{(\nu)}(\x) g_{d}(\x)\big] $ has full row rank for all $\x \in \mathcal{C} \cap \mathcal{M}$.
\end{assumption}

{\spaceskip=.9\fontdimen2\font plus \fontdimen3\font minus \fontdimen4\font Full row rank ensures compatibility solvability through the algebraic-state rates and input; the guarantees apply only where it holds on $\mathcal{C}\cap\mathcal{M}$ and to the exact model on $\mathcal{M}$, excluding rank loss, plant mismatch, and off-manifold perturbations.\par}

Under Assumption~\ref{assumption:regularity}, the constrained dynamics on $\mathcal{M}$ admit the control-affine form $\dot{\x} = \hat{f}(\x) + \hat{g}(\x)\ci$, where the \emph{projected vector fields} are
\begin{equation*}
    \hat{f}(\x) \!:=\!
        \begin{bmatrix}
            f_{d} \\
            -(J_{a}^{(1)})^{\dagger} J_{d}^{(1)} f_{d}\\
            \vdots \\
            -(J_{a}^{(\nu)})^{\dagger} J_{d}^{(\nu)} f_{d}
        \end{bmatrix}\!,\;
    \hat{g}(\x) \!:=\!
        \begin{bmatrix}
            g_{d} \\
            -(J_{a}^{(1)})^{\dagger} J_{d}^{(1)} g_{d}\\
            \vdots \\
            -(J_{a}^{(\nu)})^{\dagger} J_{d}^{(\nu)} g_{d}
        \end{bmatrix}\!,
\end{equation*}
where all entries depend on $\x$ (suppressed for brevity). For index-1 systems, only the $k=1$ row appears. The projection operator at level $k$ is $P^{(k)}(\x) := I_{n_m} - J_{a}^{(k)}(\x) (J_{a}^{(k)}(\x))^{\dagger}$, and a control $\ci$ is \emph{compatible} if it satisfies the hierarchy of compatibility conditions for all $k \in \{1,\ldots,\nu\}$:
\begin{equation}\label{eq:compatibility}
P^{(k)}(\x) \big(J_{d}^{(k)}(\x) f_{d}(\x) + \eta^{(k)}(\x) g_{d}(\x) \ci\big) = 0.
\end{equation}
Let $\mathcal{L}_{\hat{f}} \psi(\x) := \nabla \psi(\x)^{\top} \hat{f}(\x)$ and $\mathcal{L}_{\hat{g}} \psi(\x) := \nabla \psi(\x)^{\top} \hat{g}(\x)$ for the Lie derivatives along the projected fields. For notational convenience, we use the unified bar notation: $\bar{P}(\x)$, $\bar{J}_{d}(\x)$, and $\bar{\eta}(\x)$ denote $P^{(1)}(\x)$, $J_{d}^{(1)}(\x)$, $\eta^{(1)}(\x)$ for index-1, or the stacked quantities $P^{(k)}(\x)$, $J_{d}^{(k)}(\x)$, $\eta^{(k)}(\x)$ for all $k \in \{1,\ldots,\nu\}$ for index-$\nu$ systems.

Given a candidate barrier function $b(\x)$ and $\alpha(\x)$, the \emph{DAE-aware CBF-QP} selects the minimally invasive safe control:
\begin{equation}\label{eq:dae_cbf_qp}
\begin{aligned}
    \min_{\ci} \quad & \|\ci - \ci_{\mathrm{nom}}\|^2 \\
    \mathrm{s.t.} \quad & \mathcal{L}_{\hat{f}} b(\x) + \alpha(b(\x)) + \mathcal{L}_{\hat{g}} b(\x)\,\ci \ge 0, \\
    & \bar{P}(\x)\big(\bar{J}_d(\x) f_d(\x) + \bar{\eta}(\x) g_d(\x)\,\ci\big) = 0, \\
    & A_u \ci \le r_u,
\end{aligned}
\end{equation}
{\spaceskip=.9\fontdimen2\font plus \fontdimen3\font minus \fontdimen4\font where $\ci_{\mathrm{nom}}$ is a nominal controller. The first constraint enforces the CBF inequality along the projected dynamics, the second compatibility~\eqref{eq:compatibility}, and the third input bounds.}
\begin{definition}[DAE-aware CBF~\cite{zhang2026verification}]\label{def:dae_aware_cbf}
A continuously differentiable function $b(\x)$ on $\mathcal{M}$ is a \emph{DAE-aware CBF} if for every $\x \in \mathcal{D} \cap \mathcal{M}$, the QP~\eqref{eq:dae_cbf_qp} is feasible.
\end{definition}

A DAE-aware CBF $b(\x)$ is \emph{valid} for $\mathcal{C}$ if it additionally satisfies correctness, $\mathcal{D}\cap\mathcal{M}\subseteq\mathcal{C}\cap\mathcal{M}$. Under these conditions, the following theorem guarantees safety.

\begin{theorem}[Safety of DAE systems~\cite{zhang2026verification}]\label{thm:dae_safety}
Let $b(\x)$ be a valid DAE-aware CBF for $\mathcal{C}$. Then $\mathcal{D} \cap \mathcal{M}$ is positively invariant under any control policy solving~\eqref{eq:dae_cbf_qp}.
\end{theorem}



{\spaceskip=.95\fontdimen2\font plus \fontdimen3\font minus \fontdimen4\font Validity on a DAE requires that the CBF is jointly compatible with the manifold geometry. The positive invariance guarantee of Theorem~\ref{thm:dae_safety} relies on the feasibility of~\eqref{eq:dae_cbf_qp}. Appending~\eqref{eq:compatibility} to an existing ODE-based CBF does not, in general, preserve feasibility, making barrier \emph{synthesis} essential.\par}

\subsection{Problem Statement}\label{subsec:problem-statement}
Theorem~\ref{thm:dae_safety} guarantees safety of the DAE system~\eqref{eq:dyn} provided a valid DAE-aware CBF (Definition~\ref{def:dae_aware_cbf}) exists. The main problem addressed in this paper is to synthesize such a valid DAE-aware CBF, which can be written as follows.

\begin{problem}[DAE CBC Synthesis]\label{problem:dae_cbf_synthesis}
Given~\eqref{eq:dyn}, $\mathcal{C}$, and a barrier class $\mathcal{B}=\{b(\cdot;\theta):\theta\in\Theta\}$ with parameter set $\Theta$, find $\theta\in\Theta$, certificate polynomials $\Xi$, and a certificate margin $\gamma>0$ such that $b(\cdot;\theta)$ is a valid DAE-aware CBF. The tuple $(b,\Xi,\gamma)$ is the synthesized CBC.
\end{problem}

{\spaceskip=.9\fontdimen2\font plus \fontdimen3\font minus \fontdimen4\font The synthesis must produce three tightly coupled objects: a barrier $b(\x)$ defining the safe set $\mathcal{D}$, certificates certifying that a compatible control input exists at every $\x \in \mathcal{D} \cap \mathcal{M}$, and the resulting safe control policy.\par}

\section{CBCs for Differential-Algebraic Systems}\label{sec:cbc_theory}
Solving Problem~\ref{problem:dae_cbf_synthesis} requires constructing a barrier $b(\x)$ and certificate polynomials $\Xi$ that jointly certify correctness ($\mathcal{D} \cap \mathcal{M} \subseteq \mathcal{C} \cap \mathcal{M}$) and feasibility (a compatible control input exists at every $\x \in \mathcal{D} \cap \mathcal{M}$). The barrier and certificates are coupled bilinearly, so they cannot be optimized in a single convex program.
The correctness guarantee is built into the barrier parameterization itself (the \emph{anchored} formulation), while feasibility is enforced through sum-of-squares (SOS) certificates. This section develops these components and the full synthesis algorithm is presented in Section~\ref{section:synthesis}.


\parstart{Algebraic preliminaries} We recall semi-algebraic tools.
\begin{lemma}[Farkas' Lemma~\cite{matousek2006understanding}]\label{Lemma:Farkas}
Let $A \in \mathbb{R}^{m \times n}$ and $r \in \mathbb{R}^{m}$. Exactly one of the following holds: (i)~$A\ci \le r$ has a solution $\ci$; or (ii)~there exists $\lambda \ge 0$ with $\lambda^{\top} A = 0^{\top}$ and $\lambda^{\top} r < 0$.
\end{lemma}

For polynomial set emptiness, we use the Positivstellensatz. Let $\Sigma[\x]$ denote the cone of SOS polynomials, $\mathcal{N}[\cdot]$ the monoid, and $\mathbb{I}[\cdot]$ the ideal generated by polynomial families.

\begin{theorem}[Positivstellensatz~\cite{parrilo2003semidefinite}]\label{th:Positivstellensatz}
Let $(\Phi_{i})$, $(\chi_{j})$, $(\Gamma_{\ell})$ be finite families of polynomials. The set $\{\x : \Phi_{i}(\x) \ge 0,\; \chi_{j}(\x) \neq 0,\; \Gamma_{\ell}(\x) = 0\}$ is empty if and only if there exist $\Phi \in \Sigma$, $\chi \in \mathcal{N}$, $\Gamma \in \mathbb{I}$ such that $\Phi + \chi^{2} + \Gamma = 0$.
\end{theorem}

\parstart{Anchored SOS formulation} The candidate barrier function $b(\x)$ is \emph{anchored} to the safe set function $h(\x)$ through the following polynomial representation:
\begin{equation}
    b(\x) = h(\x) - s(\x)^{2} - \mu\, \phi(\x)^{\top} \phi(\x),
    \label{eq:anchored_b_form}
\end{equation}
where $s(\x)^{2} := m_{s}(\x)^{\top} Q_{s}\, m_{s}(\x)$ with $m_{s}(\x)$ a monomial basis of degree~$d_{s}$ and $Q_{s} \succeq 0$ a positive semidefinite (PSD) Gram matrix (the decision variable), and $\mu > 0$ is a fixed scalar hyperparameter. The PSD constraint ensures $s(\x)^{2} \ge 0$ everywhere, while $\mu\,\|\phi(\x)\|^{2} \ge 0$ by construction. It enforces
$$b(\x) \le h(\x)\quad \forall \x,
\qquad
b(\x)=h(\x)-s(\x)^{2}\quad \text{on } \mathcal{M},$$
which ensures global correctness $\mathcal{D}\subseteq \mathcal{C}$. The term $\mu\,\phi^{\top}\!\phi$ concentrates $\mathcal{D}$ near $\mathcal{M}$ but vanishes on $\mathcal{M}$, so it does not affect the Lie derivatives or feasibility conditions restricted to $\mathcal{M}$. Crucially, $b(\x)$ is \emph{linear} in the entries of $Q_{s}$ (for fixed $\mathbf{x}$), keeping both alternation steps as SDPs. The special choice $Q_s=0$ gives $b=h$ on $\mathcal{M}$ and is valid when the compatibility-constrained CBF-QP is feasible on $\mathcal{C}\cap\mathcal{M}$; otherwise, the parameterization searches for a smaller $\mathcal{D}\subseteq\mathcal{C}$. It remains to certify feasibility through the SOS constraints.

\parstart{Dual feasibility conditions}
For any $\x \in \mathcal{D} \cap \mathcal{M}$, the DAE-aware CBF conditions require a control $\ci \in \mathcal{U}$ satisfying compatibility, the CBF inequality, and input limits:
\begin{equation}
\begin{aligned}
    \bar{P}(\x) \big(\bar{J}_{d}(\x) f_{d}(\x) + \bar{\eta}(\x) g_{d}(\x) \ci\big) &= 0, \\
    \mathcal{L}_{\hat{f}} b(\x) + \alpha\big(b(\x)\big)
        + \mathcal{L}_{\hat{g}} b(\x) \, \ci &\geq 0, \\
    A_{u} \ci &\leq r_{u}.
\end{aligned}
    \label{eq:dae_sufficient_conditions}
\end{equation}
Encode the compatibility equality by two inequalities and write the resulting input constraints as $A(\x)\ci\le r(\x)$. For $\x\in\operatorname{int}(\mathcal{D})\cap\mathcal{M}$, $(A_{\mathrm{int}},r_{\mathrm{int}})$ contains compatibility and input bounds. For $\x\in\partial\mathcal{D}\cap\mathcal{M}$, Nagumo's theorem~\cite{blanchini2008set} adds the tangency row, giving $A_{\mathrm{bnd}}=[A_{\mathrm{int}};{-}\mathcal{L}_{\hat g}b]$ and $r_{\mathrm{bnd}}=[r_{\mathrm{int}};\mathcal{L}_{\hat f}b]$ because $\alpha(b)=0$ on the boundary.

By Lemma~\ref{Lemma:Farkas}, feasibility of each system is equivalent to infeasibility of its dual:
\begin{align}
    \lambda^{\top} A_{\mathrm{int}}(\x) = 0^{\top},
    &\qquad \lambda^{\top} r_{\mathrm{int}}(\x) < 0,
        \label{eq:dae_dual_int} \\
    \lambda^{\top} A_{\mathrm{bnd}}(\x) = 0^{\top},
    &\qquad \lambda^{\top} r_{\mathrm{bnd}}(\x) < 0,
        \label{eq:dae_dual_bnd}
\end{align}
with nonnegative $\lambda = [(\lambda^{-})^{\top}, (\lambda^{+})^{\top}, \lambda^{b}, (\lambda^{u})^{\top}]^{\top}$ (the multiplier $\lambda^{b}$ is absent in the interior dual).
To apply the Positivstellensatz to these dual systems, we impose the following representation condition.





\begin{assumption}\label{assumption:polynomial}
$h$ and $\phi$ are polynomial, and the entries of $A_{\mathrm{int}}$, $r_{\mathrm{int}}$, $A_{\mathrm{bnd}}$, and $r_{\mathrm{bnd}}$ admit polynomial representations on $\mathcal{C}\cap\mathcal{M}$.
\end{assumption}

Rational entries with denominators nonzero on $\mathcal{C}\cap\mathcal{M}$ may instead be converted by exact polynomial lifting before forming the certificates.

\parstart{Positivstellensatz certificates}
For a given $b(\x)$, feasibility requires that the Farkas dual sets~\eqref{eq:dae_dual_int}-\eqref{eq:dae_dual_bnd} are empty for all $\x \in \mathcal{D} \cap \mathcal{M}$. We certify emptiness via the Positivstellensatz (Theorem~\ref{th:Positivstellensatz}): for each $\star \in \{\mathrm{int},\mathrm{bnd}\}$, emptiness holds if there exist a \emph{cone} element $\Phi_{\star} \in \Sigma$, a \emph{monoid} element $\chi_{\star}^{2}$ (with $\chi_{\star} \in \Sigma$), and an \emph{ideal} element $\varphi_{\star} \in \mathbb{I}$ satisfying $\Phi_{\star} + \chi_{\star}^{2} + \varphi_{\star} = 0$.

The \emph{cone} encodes the nonnegative quantities: multiplier nonnegativity ($[\lambda_{\star}]_{i} \ge 0$) and, in the interior, barrier positivity ($b(\x) \ge 0$):
\begin{equation}\label{eq:sos_int_cone}
\Phi_{\mathrm{int}} := \sigma^{0}_{\mathrm{int}} + \sigma^{b}_{\mathrm{int}}\, b(\x) + \textstyle\sum_{i} \sigma^{\lambda,i}_{\mathrm{int}}\, [\lambda_{\mathrm{int}}]_{i},
\end{equation}
with all $\sigma$-multipliers in $\Sigma[\x,\lambda_{\mathrm{int}}]$. The boundary cone $\Phi_{\mathrm{bnd}}$ has the same form but \emph{omits} the $\sigma^{b} b$ term, since $b = 0$ on $\partial\mathcal{D}$; instead, $b(\x)$ enters the boundary ideal.

The \emph{ideal} encodes the equality constraints, Farkas stationarity ($\lambda_{\star}^{\top} A_{\star} = 0^{\top}$), the manifold ($\phi = 0$), and the strict-inequality slack ($-\lambda_{\star}^{\top} r_{\star} - \chi_{\star}^{2} - \gamma = 0$, where $\gamma \ge 0$ is the certificate margin):
\begin{equation*}
\varphi_{\star} := (\rho^{A}_{\star})^{\top}\!(\lambda_{\star}^{\top} A_{\star}) + (\boldsymbol{\rho}^{\phi}_{\star})^{\top}\!\phi + \rho^{s}_{\star} ({-}\lambda_{\star}^{\top} r_{\star} {-} \chi_{\star}^{2} {-} \gamma),
\end{equation*}
with an additional term $\rho^{b}_{\mathrm{bnd}}\, b(\x)$ appended for $\star = \mathrm{bnd}$ (since $b = 0$ on $\partial\mathcal{D}$). All $\rho$ are free polynomial multipliers. Let $\Xi:=\{\sigma_{\star},\chi_{\star},\rho_{\star}\}_{\star\in\{\mathrm{int},\mathrm{bnd}\}}$ collect the certificate polynomials, and let $\gamma>0$ denote the strict certificate margin. The full SOS certificates are:
\begin{subequations}
\begin{align}
\Phi_{\mathrm{int}} + \chi_{\mathrm{int}}^{2} + \varphi_{\mathrm{int}} &= 0, \label{eq:sos_int_cert} \\
\Phi_{\mathrm{bnd}} + \chi_{\mathrm{bnd}}^{2} + \varphi_{\mathrm{bnd}} &= 0. \label{eq:sos_bnd_cert}
\end{align}
\end{subequations}

The constraints in \eqref{eq:sos_int_cone}-\eqref{eq:sos_bnd_cert} are bilinear in the unknown barrier and certificate polynomials, motivating the alternating scheme of Section~\ref{section:synthesis}. The following theorem establishes that the anchored formulation together with the Psatz certificates constitutes a valid control barrier certificate.

\begin{figure}[t]
\vspace*{1mm}
\centering
\definecolor{ink}{RGB}{18,18,18}
\definecolor{gfill}{RGB}{248,248,248}
\definecolor{gsafe}{RGB}{232,232,232}
\definecolor{gunsafe}{RGB}{205,205,205}
\definecolor{gmfld}{RGB}{145,145,145}
\definecolor{gbnd}{RGB}{88,88,88}
\tikzset{
  panelborder/.style={ink, line width=0.80pt},
}

\def\Pw{5.50}    
\def\Ph{2.30}    
\def\Gap{0.10}   
\def\Lw{1.50}    

\resizebox{0.72\columnwidth}{!}{%
\begin{tikzpicture}[font=\rmfamily, >=Latex, line cap=round, line join=round]

\begin{scope}[shift={(0,\Ph+\Gap)}]
  \fill[gfill] (0,0) rectangle (\Lw,\Ph);
  \draw[panelborder, rounded corners=1.5pt] (0,0) rectangle (\Lw,\Ph);
  \node[font=\bfseries\fontsize{8}{9}\selectfont, text=ink, anchor=north east] at (-0.08,\Ph) {(a)};
  \node[font=\bfseries\fontsize{9}{10}\selectfont, text=ink] at (0.75,1.55) {ODE};
  \node[font=\fontsize{6.5}{8}\selectfont, text=ink, align=center] at (0.75,1.22) {standard\\$b=h-s^{2}$};

  \draw[->, ink, line width=0.45pt] (0.24,0.23) -- (1.26,0.23);
  \draw[->, ink, line width=0.45pt] (0.38,0.13) -- (0.38,0.95);

  \draw[ink, line width=1.1pt]
    (0.30,0.15) .. controls (0.55,0.71) and (0.80,0.63) ..
    (1.05,0.44) .. controls (1.14,0.36) and (1.20,0.40) .. (1.30,0.42);
\end{scope}

\begin{scope}[shift={(\Lw+\Gap, \Ph+\Gap)}]
  \fill[gsafe] (0,0) rectangle (\Pw,\Ph);
  \draw[panelborder] (0,0) rectangle (\Pw,\Ph);

  \draw[gbnd, line width=1.6pt]
    (4.21,0.00) .. controls (3.77,0.65) and (3.70,1.29) .. (4.58,\Ph);

  \fill[gunsafe]
    (4.21,0.00) .. controls (3.77,0.65) and (3.70,1.29) .. (4.58,\Ph)
    -- (\Pw,\Ph) -- (\Pw,0) -- cycle;
  \draw[gbnd, line width=1.6pt]
    (4.21,0.00) .. controls (3.77,0.65) and (3.70,1.29) .. (4.58,\Ph);

  \draw[ink, line width=1.3pt]
    (3.47,0.00) .. controls (3.10,0.65) and (3.03,1.29) .. (3.81,\Ph);

  \draw[<->, ink, line width=0.55pt] (3.28,1.13) -- (3.96,1.13);
  \node[font=\fontsize{6}{7}\selectfont, text=ink, above] at (3.62,1.15) {$s^{2}$};

  \node[font=\fontsize{6.5}{8}\selectfont, text=ink] at (0.85,0.18)
    {$\mathcal{C}$: $h\!\ge\!0$};
  \node[font=\fontsize{6.5}{8}\selectfont, text=ink] at (1.30,1.45)
    {$\mathcal{D}\!=\!\mathcal{C}$ shrunk by $s^{2}$};
  \node[font=\fontsize{6.5}{8}\selectfont, text=ink, anchor=west] at (4.70,0.18)
    {$h\!<\!0$};
  \node[font=\fontsize{6.5}{8}\selectfont, text=ink, anchor=east] at (3.40,0.18)
    {$b\!\ge\!0$};

  \node[font=\fontsize{6}{7}\selectfont, text=ink, anchor=east] at (3.08,0.77)
    {$b\!=\!0$};
  \node[font=\fontsize{6}{7}\selectfont, text=ink, anchor=west] at (4.02,0.77)
    {$h\!=\!0$};

  \node[font=\fontsize{5.8}{7}\selectfont, text=ink, align=center] at (2.95,1.74)
    {shrinkage};
\end{scope}

\begin{scope}[shift={(0,0)}]
  \fill[gfill] (0,0) rectangle (\Lw,\Ph);
  \draw[panelborder, rounded corners=1.5pt] (0,0) rectangle (\Lw,\Ph);
  \node[font=\bfseries\fontsize{8}{9}\selectfont, text=ink, anchor=north east] at (-0.08,\Ph) {(b)};
  \node[font=\bfseries\fontsize{9}{10}\selectfont, text=ink] at (0.75,1.70) {DAE};
  \node[font=\fontsize{6}{7}\selectfont, text=ink, align=center] at (0.75,1.22)
    {anchored\\$b=h{-}s^{2}$\\${-}\mu\,\phi^{\!\top}\!\phi$};

  \draw[->, ink, line width=0.45pt] (0.24,0.23) -- (1.26,0.23);
  \draw[->, ink, line width=0.45pt] (0.38,0.13) -- (0.38,0.95);

  \draw[gmfld, line width=0.75pt]
    (0.30,0.36) .. controls (0.58,0.74) and (0.92,0.73) .. (1.28,0.58);
  \draw[gmfld, line width=0.75pt]
    (0.34,0.18) .. controls (0.64,0.47) and (0.98,0.47) .. (1.30,0.37);

  \draw[ink, line width=1.1pt]
    (0.32,0.24) .. controls (0.58,0.58) and (0.80,0.57) ..
    (1.05,0.44) .. controls (1.14,0.39) and (1.20,0.40) .. (1.28,0.42);
\end{scope}

\begin{scope}[shift={(\Lw+\Gap, 0)}]
  \fill[gsafe] (0,0) rectangle (\Pw,\Ph);
  \draw[panelborder] (0,0) rectangle (\Pw,\Ph);

  \fill[gunsafe]
    (4.21,0.00) .. controls (3.77,0.65) and (3.70,1.29) .. (4.58,\Ph)
    -- (\Pw,\Ph) -- (\Pw,0) -- cycle;
  \draw[gbnd, line width=1.6pt]
    (4.21,0.00) .. controls (3.77,0.65) and (3.70,1.29) .. (4.58,\Ph);

  \begin{scope}
    \clip (0.00,0.00) -- (3.47,0.00)
      .. controls (3.10,0.65) and (3.03,1.29) .. (3.81,\Ph)
      -- (0.00,\Ph) -- cycle;
    \fill[ink, opacity=0.10]
      (0.00,1.01) .. controls (0.74,1.39) and (1.63,1.57) ..
      (2.36,1.49) .. controls (2.96,1.44) and (3.55,1.52) ..
      (4.14,1.74) .. controls (4.51,1.87) and (4.80,2.06) .. (5.17,\Ph)
      -- (\Pw,1.90)
      .. controls (5.03,1.65) and (4.58,1.36) ..
      (4.14,1.17) .. controls (3.55,0.90) and (2.96,0.79) ..
      (2.36,0.85) .. controls (1.63,0.93) and (0.74,0.71) ..
      (0.00,0.23)
      -- cycle;
  \end{scope}

  \draw[ink, line width=1.3pt]
    (3.47,0.00) .. controls (3.10,0.65) and (3.03,1.29) .. (3.81,\Ph);

  \draw[gmfld, line width=1.8pt]
    (0.00,0.61) .. controls (0.74,1.05) and (1.63,1.25) ..
    (2.36,1.17) .. controls (2.96,1.11) and (3.55,1.21) ..
    (4.14,1.45) .. controls (4.58,1.63) and (4.95,1.90) .. (\Pw,2.14);

  \draw[ink, line width=0.9pt, dashed]
    (0.00,1.01) .. controls (0.74,1.39) and (1.63,1.57) ..
    (2.36,1.49) .. controls (2.96,1.44) and (3.55,1.52) ..
    (4.14,1.74) .. controls (4.51,1.87) and (4.80,2.06) .. (5.17,\Ph);

  \draw[ink, line width=0.9pt, dashed]
    (0.00,0.23) .. controls (0.74,0.71) and (1.63,0.93) ..
    (2.36,0.85) .. controls (2.96,0.79) and (3.55,0.90) ..
    (4.14,1.17) .. controls (4.58,1.36) and (5.03,1.65) .. (\Pw,1.90);

  \draw[<->, ink, line width=0.55pt] (3.33,0.32) -- (4.01,0.32);
  \node[font=\fontsize{6}{7}\selectfont, text=ink, above] at (3.68,0.34) {$s^{2}$};

  \draw[<->, ink, line width=0.45pt]
    (1.63,1.21) -- (1.63,1.53);
  \node[font=\fontsize{5.5}{6.5}\selectfont, text=ink, anchor=west] at (1.68,1.37)
    {$\mu\|\phi\|^{2}$};
  \draw[<->, ink, line width=0.45pt]
    (1.63,0.89) -- (1.63,1.21);
  \node[font=\fontsize{5.5}{6.5}\selectfont, text=ink, anchor=west] at (1.68,1.05)
    {$\mu\|\phi\|^{2}$};

  \node[font=\fontsize{6.5}{8}\selectfont, text=ink] at (0.60,0.18)
    {$\mathcal{C}$: $h\!\ge\!0$};
  \node[font=\fontsize{6.5}{8}\selectfont, text=ink, anchor=west] at (4.70,0.18)
    {$h\!<\!0$};

  \node[font=\fontsize{6.5}{8}\selectfont, text=ink, anchor=east] at (1.80,1.86)
    {$\mathcal{M}$: $\phi\!=\!0$};
  \draw[->, ink, line width=0.5pt] (1.22,1.74) .. controls (1.22,1.65) .. (1.22,1.09);

\end{scope}

\end{tikzpicture}%
}

\caption{(a) Standard ODE barrier: the safe set $\mathcal{C}$ is shrunk by $s^{2}$ to yield the barrier set $\mathcal{D} \subseteq \mathcal{C}$. (b) DAE anchored barrier: the off-manifold penalty $\mu\|\phi\|^{2}$ vanishes on $\mathcal{M}$ and grows quadratically away from it, concentrating $\mathcal{D}$ near the constraint manifold while preserving $b \le h$ everywhere.}
\label{fig:anchored_geometry}
\end{figure}
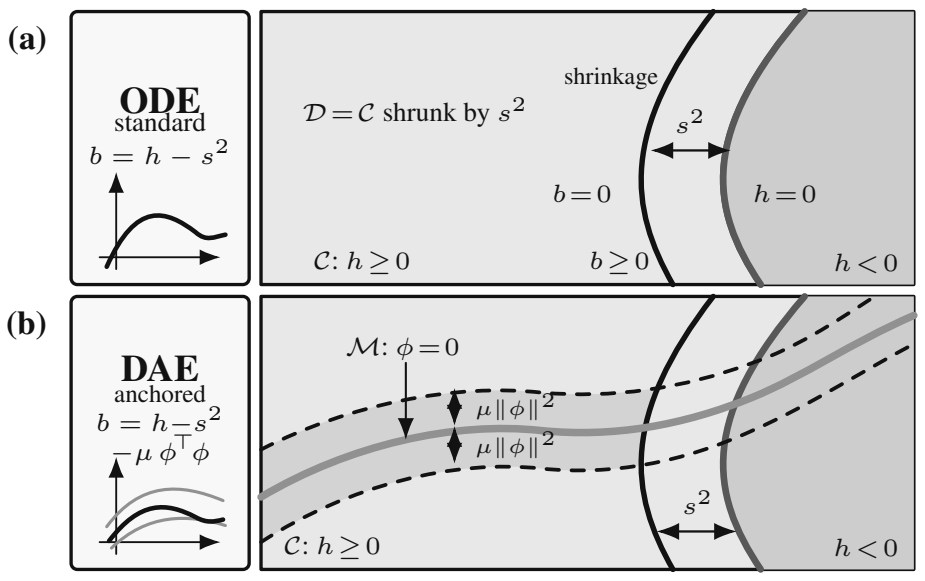

\begin{theorem}[Control barrier certificate for DAEs]\label{th:cbcdae}
Under Assumption~\ref{assumption:polynomial}, let $b(\x)$ be defined by~\eqref{eq:anchored_b_form} with $Q_s \succeq 0$ and $\mu > 0$. If there exist certificate polynomials $\Xi$ and margin $\gamma > 0$ satisfying~\eqref{eq:sos_int_cert}-\eqref{eq:sos_bnd_cert}, then $b(\x)$ is a control barrier certificate for DAE system \eqref{eq:dyn}, i.e.,
\begin{subequations}\label{eq:cbc_conditions}
\begin{align}
    \label{eq:safe_cbc}
    b(\x) &\geq 0, \quad \forall \x \in \mathcal{D} \cap \mathcal{M}, \\
    \label{eq:unsafe_cbc}
    b(\x) &< 0, \quad \forall \x \in (\mathcal{X}\setminus\mathcal{C}) \cap \mathcal{M},  \\
    \label{eq:feasible_cbc}
    \exists \alpha, \ci &\in \mathcal{U}, \, \text{s.t.} \, \eqref{eq:dae_sufficient_conditions} \,\text{holds},\quad \forall \x \in \mathcal{D} \cap \mathcal{M},
\end{align}
\end{subequations}
and $\mathcal{D} \cap \mathcal{M}$ is positively invariant under policy~\eqref{eq:dae_cbf_qp}.
\end{theorem}
\begin{proof}
    Given anchored form~\eqref{eq:anchored_b_form} with $Q_s \succeq 0$ and $\mu > 0$, we have $s(\x)^{2}$ and $\mu\phi(\x)^{T}\phi(\x)$ nonnegative everywhere. It gives us $\mathcal{D}\subseteq \mathcal{C}$ and $\mathcal{D}\cap\mathcal{M}$ is nonempty, if $\mathcal{D}$ is nonempty, respectively. Hence, it enforces correctness~\eqref{eq:safe_cbc}-\eqref{eq:unsafe_cbc} by construction. 
    For each $\star \in \{\mathrm{int}, \mathrm{bnd}\}$, the Positivstellensatz conditions~\eqref{eq:sos_int_cert}--\eqref{eq:sos_bnd_cert} with $\gamma > 0$ certifies that the Farkas dual set~\eqref{eq:dae_dual_int}-\eqref{eq:dae_dual_bnd} is empty for every $\x \in \mathcal{D} \cap \mathcal{M}$. By Lemma~\ref{Lemma:Farkas}, the primal system~\eqref{eq:dae_sufficient_conditions} therefore admits a solution $\ci \in \mathcal{U}$, establishing~\eqref{eq:feasible_cbc}.
    Since $b(\x)$ is a valid DAE-aware CBF (Definition~\ref{def:dae_aware_cbf}), $\mathcal{D} \cap \mathcal{M}$ is positively invariant under any policy solving~\eqref{eq:dae_cbf_qp} by Theorem~\ref{thm:dae_safety}.
\end{proof}
{\spaceskip=.95\fontdimen2\font plus \fontdimen3\font minus \fontdimen4\font Theorem~\ref{th:cbcdae} reduces the CBC synthesis problem to finding a Gram matrix $Q_s \succeq 0$ and certificate polynomials $\Xi$ that jointly satisfy the Psatz identities~\eqref{eq:sos_int_cert}-\eqref{eq:sos_bnd_cert} with $\gamma > 0$. Since $(Q_s, \Xi)$ appear bilinearly in these identities, a single convex program cannot optimize both simultaneously. Section~\ref{section:synthesis} presents an alternating SOS algorithm (Algorithm~\ref{alg:alternating_sos_cbf}) that decouples the search into convex subproblems, together with a Lyapunov-based initialization that provides a certified warm start.\par}


\section{Alternating SOS Synthesis}\label{section:synthesis}
\begingroup
\spaceskip=.9\fontdimen2\font plus \fontdimen3\font minus \fontdimen4\font
We first construct a certified initial barrier via a Lyapunov-based SDP (Proposition~\ref{prop:init}), then present the two alternating convex steps and their assembly into Algorithm~\ref{alg:alternating_sos_cbf}.
The control-influence matrix $\eta^{(k)}$ accounts for higher relative degree in DAE compatibility. The SOS program presents the relative-degree-one barrier condition; the higher-order case is omitted.\par
\endgroup


\parstart{Initialization}
Let $\dot{\z}=f_{\mathrm{red}}(\z)+g_{\mathrm{red}}(\z)\ci$ denote the reduced ODE in coordinates $\z=T\x_d$. Linearizing it at an equilibrium $(\z^*,\ci^*)$ gives $A_{\mathrm{red}}:=\nabla_{\z}(f_{\mathrm{red}}+g_{\mathrm{red}}\ci^*)|_{\z^*}$ and $B_{\mathrm{red}}:=g_{\mathrm{red}}(\z^*)$. We compute the continuous-time LQR gain $K$ for $(A_{\mathrm{red}},B_{\mathrm{red}})$ from the algebraic Riccati equation; under $\ci-\ci^*=-K(\z-\z^*)$, $A_{\mathrm{cl}}=A_{\mathrm{red}}-B_{\mathrm{red}}K$. The case-study weights are reported in Section~\ref{sec:case-study}, and the Lyapunov linear matrix inequality (LMI)~\cite{boyd1993control} yields the seed $Q_s^{(0)}\succeq0$.

Let $V(\x; P) = (\x - \x^{*})^{\top} P\,(\x - \x^{*})$. We jointly optimize a Lyapunov matrix $P$, the Gram matrix $Q_{s}$, and the barrier margin $c$:
\begin{equation}
\label{eq:init_sdp}
\begin{aligned}
\max_{\,P,\,Q_{s},\,c}&\quad  c \\
\text{s.t.}\quad
  & P \succ 0,\; Q_{s} \succeq 0,\; c > 0, \; A_{\mathrm{cl}}^{\top} P + P\, A_{\mathrm{cl}} \prec 0, \\
  & m_{s}(\x)^{\top} Q_{s}\, m_{s}(\x)
    = h(\x) - c + V(\x;P)\\
  & \qquad\qquad\qquad\qquad\quad\;
    - \mu\phi(\x)^{\top}\phi(\x).
\end{aligned}
\end{equation}
All constraints are linear in $(P, Q_{s}, c)$, so~\eqref{eq:init_sdp} is a convex SDP. Since $A_{\mathrm{cl}}$ is Hurwitz, the Lyapunov LMI is always feasible~\cite{boyd1993control}, and the identity constraint links the barrier shape to a Lyapunov sublevel set. We denote the resulting initial safe set $\mathcal{D}^{(0)} := \{\x : b(\x;\, Q_{s}^{(0)}) \ge 0\}$. The following proposition is straightforward from the construction and guarantees that the initial barrier is a valid DAE-aware CBF near $\x^{*}$.

\begin{algorithm}[t]
\caption{Alternating SOS Synthesis for DAE-aware CBF}
\label{alg:alternating_sos_cbf}
\begin{algorithmic}[1]
\State \textbf{Input:} $f_{d}, g_{d}, h$, $\mu$, $\x^{*}$, polynomial degree $deg$.
\State \textbf{Init:} Given $\x_0$, synthesize $Q_{s}^{(0)}$ via~\eqref{eq:init_sdp} (Prop.~\ref{prop:init}), $k\!\leftarrow\!0$.
\Repeat
  \State $b \leftarrow$ \textsc{BuildBarrier}$(h,\phi; Q_{s}^{(k)},\mu)$. \hfill\eqref{eq:anchored_b_form}
  \State $(\Xi,\hat{\gamma}) \leftarrow$ \textsc{SolveCerts}$(b,\text{deg})$. \hfill\eqref{eq:step1_sdp}
  \State $Q_{s}^{(k+1)} \leftarrow$ \textsc{SolveBarrier}$(\Xi,\text{deg})$. \hfill\eqref{eq:step2_sdp}
  \State $k \leftarrow k + 1$.
\Until{$\|Q_{s}^{(k)} - Q_{s}^{(k-1)}\|_{F} < \varepsilon$}
\State \textbf{Output:} a verified $(Q_s,\Xi,\hat\gamma)$ if $Q_s\succeq0$, $\hat\gamma>0$, and~\eqref{eq:sos_int_cert}--\eqref{eq:sos_bnd_cert} hold after final substitution; otherwise, failure.
\end{algorithmic}
\end{algorithm}

\begin{proposition}\label{prop:init}
Given $\x^{*} \in \operatorname{int}(\mathcal{C}) \cap \mathcal{M}$, if~\eqref{eq:init_sdp} is feasible with solution $(P^{*}, Q_{s}^{(0)}, c^{*})$, then the initial barrier $b(\x; Q_{s}^{(0)})$ satisfies~\eqref{eq:cbc_conditions} with $\mathcal{U}=\mathbb{R}^{n_{u}}$.
\end{proposition}
The proof follows from the Lyapunov-based construction. The proof is omitted for brevity.
The SDP~\eqref{eq:init_sdp} does not explicitly enforce $\ci \in \mathcal{U}$. A simple two-stage remedy is to first solve~\eqref{eq:init_sdp}, then compute $\bar{c} = \min_{j} u_{\max,j}^{2} / (k_{j}^{\top} (P^{*})^{-1} k_{j})$, and finally re-solve with the additional constraint $c \le \bar{c}$.


\parstart{Step~1 (SolveCerts)} Fix $Q_{s}$ (and hence $b(\x)$ via~\eqref{eq:anchored_b_form}) and solve for the certificate polynomials $\Xi$ while maximizing the certificate margin~$\gamma$:
\begin{equation}
\label{eq:step1_sdp}
\begin{aligned}
\max_{\,\Xi,\,\gamma}\quad & \gamma \\
\text{s.t.}\quad & \text{identities } \eqref{eq:sos_int_cert} \text{ and } \eqref{eq:sos_bnd_cert} \text{ hold}, \\
& \text{all } \sigma \text{-multipliers} \in \Sigma[\x,\lambda].
\end{aligned}
\end{equation}
The Lie derivatives appearing in $A_{\mathrm{bnd}}$ and $r_{\mathrm{bnd}}$ are computed from~\eqref{eq:anchored_b_form}. The interior and boundary SDPs can be solved independently (each with its own margin $\gamma_{\mathrm{int}}, \gamma_{\mathrm{bnd}}$), and the reported margin is $\hat{\gamma} := \min\{\gamma_{\mathrm{int}}, \gamma_{\mathrm{bnd}}\}$.

\parstart{Step~2 (SolveBarrier)} Fix the certificate polynomials $\Xi$ (numerical Lagrangians from Step~1) and solve for the Gram matrix $Q_{s}$:
\begin{equation}
\label{eq:step2_sdp}
\begin{aligned}
\min_{\,Q_{s}}\quad & \|Q_{s}\|_{F}^{2} \\
\text{s.t.}\quad & \text{identities } \eqref{eq:sos_int_cert} \text{ and } \eqref{eq:sos_bnd_cert} \text{ hold with fixed } \Xi, \\
& Q_{s} \succeq 0, \quad b(\x^{*}) > 0.
\end{aligned}
\end{equation}
{\spaceskip=.9\fontdimen2\font plus \fontdimen3\font minus \fontdimen4\font The Frobenius-norm objective regularizes $Q_{s}$ toward a minimum-norm solution and pushes $b(\x)$ toward the geometrical limit $h(\x)$ while satisfying the Lagrangian constraints.\par}

We define helper routines: \textsc{BuildBarrier}$(h,\phi; Q_{s},\mu)$ returns $b(\x)$ via \eqref{eq:anchored_b_form}; \textsc{SolveCerts}$(b,\text{deg})$ solves \eqref{eq:step1_sdp} and returns $(\Xi,\hat{\gamma})$; \textsc{SolveBarrier}$(\Xi,\text{deg})$ solves \eqref{eq:step2_sdp} for $Q_{s}$.



\parstart{ODE-to-DAE lifting}
Direct SOS synthesis on high-dimensional DAEs is computationally intractable: the monomial basis grows as $\binom{n_x + d_s}{d_s}$, so increasing the state dimension from $n_{\mathrm{red}}$ to $n_x$ inflates the Gram matrix combinatorially. For example, going from $6$ to $24$ states at degree~$d_s = 2$ increases the SDP variable count by a factor of $\approx 130\times$. Instead, one synthesizes on a reduced ODE in minimal coordinates and \emph{lifts} the result to the full DAE state space.

Let $\z=T\x_d$ be a linear coordinate projection whose restriction to the considered component of $\mathcal{M}$ admits a smooth inverse embedding $\iota$, so that each consistent state is represented by $\x=\iota(\z)$. Assume that every trajectory of the reduced ODE under $\ci\in\mathcal{U}$ lifts through $\iota$ to a consistent DAE trajectory and that $h^{\mathrm{dae}}(\iota(\z))=h^{\mathrm{ode}}(\z)$ on this component. Since $T$ is linear, each ODE monomial expands as a linear combination of DAE monomials:
\begin{equation}\label{eq:monomial_embedding}
    m_s^{\mathrm{ode}}(T\x_d) = L\, m_s^{\mathrm{dae}}(\x_d),
\end{equation}
where $L \in \mathbb{R}^{\dim(m_s^{\mathrm{ode}}) \times \dim(m_s^{\mathrm{dae}})}$ is the \emph{monomial embedding matrix}. Given a Gram matrix $Q_s^{\mathrm{ode}} \succeq 0$ from Algorithm~\ref{alg:alternating_sos_cbf} on the ODE, define the lifted Gram matrix
\begin{equation}\label{eq:lifted_gram}
    Q_s^{\mathrm{dae}} := L^{\top} Q_s^{\mathrm{ode}}\, L,
\end{equation}

\begin{corollary}[ODE-to-DAE lifting]\label{thm:lifting}

Under the exact-reduction assumption, if Algorithm~\ref{alg:alternating_sos_cbf} returns a reduced-order CBC $(Q_s^{\mathrm{ode}},\Xi,\hat\gamma)$ with $Q_s^{\mathrm{ode}}\succeq0$ and $\hat\gamma>0$, then
\[
\begin{aligned}
b^{\mathrm{dae}}(\x)={}&h^{\mathrm{dae}}(\x)
-m_s^{\mathrm{ode}}(T\x_d)^\top Q_s^{\mathrm{ode}}m_s^{\mathrm{ode}}(T\x_d)\\
&-\mu\phi(\x)^\top\phi(\x)
\end{aligned}
\]
is a CBC on the considered component of $\mathcal{M}$.
\end{corollary}
\begin{proof}

On $\mathcal{M}$, $\phi=0$ and $b^{\mathrm{dae}}(\iota(\z))=b^{\mathrm{ode}}(\z)$; hence, along corresponding trajectories, $\dot b^{\mathrm{dae}}=\dot b^{\mathrm{ode}}$, and the same $\ci\in\mathcal{U}$ satisfies the DAE compatibility and CBF conditions. Substituting $\z=T\x_d$ into the certificate identities completes the proof by Theorem~\ref{th:cbcdae}.
\end{proof}

\begingroup
\spaceskip=.95\fontdimen2\font plus \fontdimen3\font minus \fontdimen4\font
Under its coordinate, dynamics-equivalence, and safety-function assumptions, Corollary~\ref{thm:lifting} establishes that a reduced-order CBC returned by Algorithm~\ref{alg:alternating_sos_cbf} pulls back to a valid CBC on the considered component of $\mathcal{M}$. This avoids the combinatorial growth of the SDP variables with state dimension. We note that Algorithm~\ref{alg:alternating_sos_cbf} has no claimed convergence or global-optimality guarantee. A returned tuple is a CBC only if the final $(Q_s,\Xi,\hat\gamma)$ satisfies the Positivstellensatz identities with $Q_s\succeq0$ and $\hat\gamma>0$; otherwise, the algorithm returns no certificate and makes no safety claim. Furthermore, Algorithm~\ref{alg:alternating_sos_cbf} can be further accelerated by replacing every SOS constraint with its diagonally dominant SOS (DSOS) relaxation. Concretely, the Positive Semi-Definite (PSD) constraint $Q_{s} \succeq 0$ is replaced by diagonal dominance ($[Q_{s}]_{ii} \ge \sum_{j \neq i} |[Q_{s}]_{ij}|$), and each SOS cone multiplier $\sigma \in \Sigma$ is restricted to have a diagonally dominant Gram matrix. Since diagonal dominance implies PSD, any returned DSOS certificate remains sound under Theorem~\ref{th:cbcdae}. DSOS can be more conservative than SOS, and we make no safe-set optimality claim.\par
\endgroup

\begin{figure*}[ht]
\centering
\includegraphics[width=0.66\textwidth]{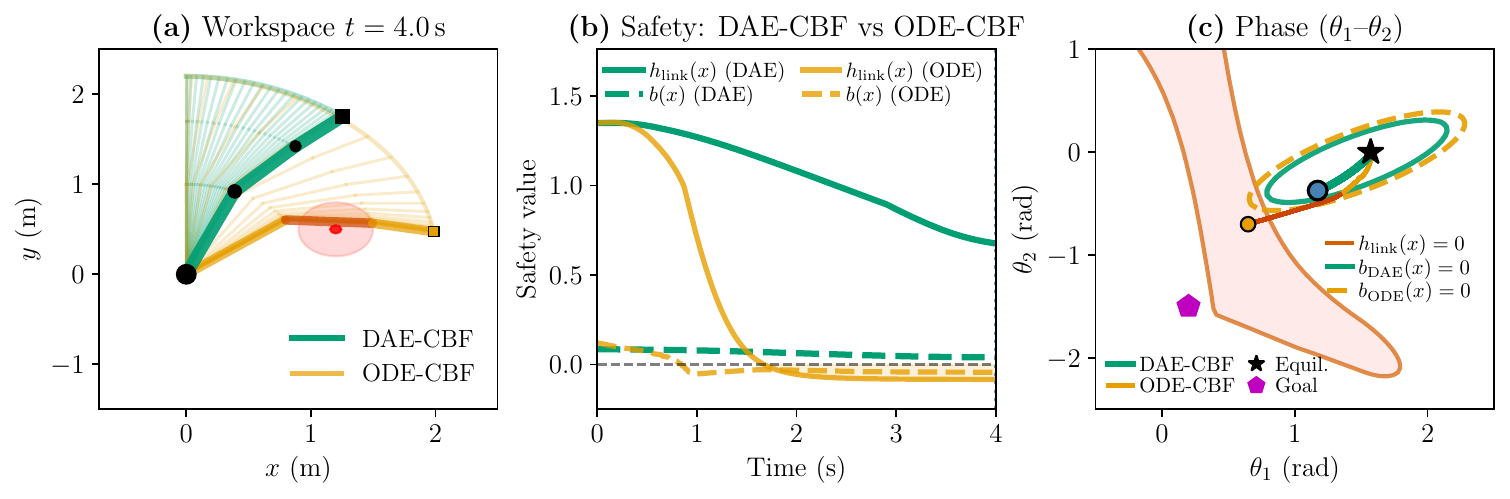}
\vspace{-0.3cm}
\caption{Closed-loop simulation of the three-link planar comparing DAE-aware CBC (green) and ODE-only CBC (orange). (a)~Workspace at $t = 4.0$\,s: the DAE-aware CBF keeps all links avoiding the obstacle (red circle), while the ODE-only CBF results in collision. (b)~Safety function $h_{\mathrm{link}}$~\eqref{eq:h_wb} and barrier $b(\x)$ over time: both remain nonnegative under DAE-CBC, whereas $h_{\mathrm{link}}$ drops to zero under ODE-CBC. (c)~Phase portrait in $(\theta_{1}, \theta_{2})$: the collision boundary ($h_{\mathrm{link}} = 0$, red), the DAE barrier zero-level set ($b_{\mathrm{DAE}} = 0$, green), the ODE barrier zero-level set ($b_{\mathrm{ODE}} = 0$, orange). The DAE-CBC trajectory remains inside $\{b_{\mathrm{DAE}} \ge 0\} \subset \{h_{\mathrm{link}} \ge 0\}$, while the ODE-CBC trajectory exits the safe region. 
}
\label{fig:three_link_sim}
\vspace{-0.2in}
\end{figure*}

\section{Case Study}\label{sec:case-study}
\begingroup
\spaceskip=.9\fontdimen2\font plus \fontdimen3\font minus \fontdimen4\font
We validate the proposed framework on a planar three-link manipulator~\cite{lai2016stable} operating near an obstacle, and we pose the following research questions.\par
\endgroup
\begin{enumerate}[label=\textbf{RQ\arabic*},leftmargin=*,itemsep=1pt]
    \item\label{rq:safety} Does the proposed DAE-aware CBC ensure safety where an ODE-CBC fails?
    \item\label{rq:scalability} How scalable is the proposed synthesis algorithm?
\end{enumerate}
The experiments are conducted in a simulation environment using Python and are performed on a machine with an Apple M2 Max with 32 GB RAM running macOS. The solver for the SOS programs is Drake~\cite{drake} and the code is available\footnote{{\url{https://github.com/HongchaoZhang-HZ/CBCDAE}}}.

\subsection{System Model}\label{subsec:setup}

\begingroup
\spaceskip=.9\fontdimen2\font plus \fontdimen3\font minus \fontdimen4\font
We consider a three-link manipulator with link lengths $L_1 = 1.0$, $L_2 = 0.7$, $L_3 = 0.5$\,m, masses $m_1 = 1.0$, $m_2 = 0.7$, $m_3 = 0.3$\,kg, and moments of inertia $I_i = \tfrac{1}{12}m_i L_i^2$, under gravity $g = 9.81$\,m/s$^{2}$. The base is pinned at the origin, a point obstacle lies at $p_{\mathrm{obs}} = (1.2, 0.5)$\,m with safety radius $d_{\mathrm{safe}} = 0.3$\,m, and joint torques $\ci = [\tau_1, \tau_2, \tau_3]^\top$ satisfy $|\tau_i| \le \tau_{\max}$. The equilibrium is $\theta_1^{*} = \pi/2$, $\theta_2^{*} = \theta_3^{*} = 0$. We model the system as a Newton-Euler DAE: each link has center-of-mass position $r_i = (x_i, y_i)$, orientation $\theta_i$, and velocities $(\dot{x}_i, \dot{y}_i, \omega_i)$, giving $\x_d \in \R^{18}$; the algebraic state $\x_a \in \R^{6}$ collects the pin-joint constraint forces. The dynamics are
\endgroup
\begin{subequations}\label{eq:manipulator_dae}
\begin{align}
    M_i \ddot{r}_i &= F_i^{\mathrm{ext}}(\x_a), \label{eq:manip_newton}\\
    I_i \dot{\omega}_i &= T_i^{\mathrm{ext}}(\x_a, \ci), \label{eq:manip_euler}\\
    0 &= \phi(\x_d), \label{eq:manip_constraint}
\end{align}
\end{subequations}
for $i = 1,2,3$, where $F_i^{\mathrm{ext}}$ collects gravity and constraint forces, and $T_i^{\mathrm{ext}}$ is the net torque from constraints and actuation. The six constraints~\eqref{eq:manip_constraint} enforce pin-joint connectivity: each joint coincides with the endpoints of its adjacent links via $\phi(\x_d) = 0$, where $\phi$ depends only on $\x_d$. Hence $\nabla_{\x_a}\phi = 0$, $\bar{P} = I_6$ (maximally nontrivial compatibility), and the system is index-3 with 3~true degrees of freedom.

\parstart{Safety function} The $i$-th link segment $\ell_i$ runs between $r_i \pm \tfrac{L_i}{2}e(\theta_i)$ with $e(\theta_i) = (\cos\theta_i, \sin\theta_i)^\top$. The safe set is $\mathcal{C} = \{\x : h(\x) \ge 0\}$, with whole-body safety function $h(\x)$:
\begin{equation}\label{eq:h_wb}
    h(\x) = \min_{i \in \{1,2,3\}} \;\mathrm{dist}(\ell_i, p_{\mathrm{obs}})^2 - d_{\mathrm{safe}}^2.
\end{equation}
Since~\eqref{eq:h_wb} is non-polynomial, SOS synthesis uses a degree-4 surrogate $h_{\mathrm{poly}}$ and the polynomial set $\mathcal{C}_{\mathrm{poly}}:=\{\x:h_{\mathrm{poly}}(\x)\ge0\}$. The formal CBC guarantee is $\mathcal{D}\cap\mathcal{M}\subseteq\mathcal{C}_{\mathrm{poly}}\cap\mathcal{M}$; whole-body safety with respect to~\eqref{eq:h_wb} is evaluated in simulation because $\mathcal{C}_{\mathrm{poly}}\subseteq\mathcal{C}$ is not certified here.

\parstart{Reduced ODE} The Lagrangian formulation eliminates $\x_a$, yielding a 6D ODE in $q = [\theta_1, \theta_2, \theta_3]^\top$:
\begin{equation}\label{eq:ode_manipulator}
    M(q)\ddot{q} + C(q,\dot{q})\dot{q} + G(q) = \ci,
\end{equation}
{\spaceskip=.9\fontdimen2\font plus \fontdimen3\font minus \fontdimen4\font which preserves the dynamics but discards link geometry; for SOS synthesis, the reduced and Newton-Euler representations use polynomial models in deviation coordinates and satisfy Assumption~\ref{assumption:polynomial}.\par}

\subsection{Experimental Results}\label{subsec:results}
\begingroup
\spaceskip=.9\fontdimen2\font plus \fontdimen3\font minus \fontdimen4\font
We initialize the synthesis on the DAE system~\eqref{eq:manipulator_dae} via~\eqref{eq:init_sdp}, which linearizes about the equilibrium under LQR control ($Q = \mathrm{diag}(10,1,10,1,10,1)$, $R = 0.1\,I$) and yields a Lyapunov-based seed with margin $c = 0.086$. We compare two synthesis strategies in Table~\ref{tab:sdp_scaling}: direct synthesis (DAE), and projection to the reduced ODE~\eqref{eq:ode_manipulator} followed by lifting (Red.). The anchored form~\eqref{eq:anchored_b_form} uses $h_{\mathrm{poly}}$; the monomial basis ($d_s = 2$, 6 states) gives a $28 \times 28$ Gram matrix $Q_s^{\mathrm{ode}}$. 
Algorithm~\ref{alg:alternating_sos_cbf} meets its stopping and final-certificate checks after 2~SOS iterations and 8~DSOS iterations. The ODE Gram matrix is then lifted via~\eqref{eq:lifted_gram} back to a $190 \times 190$ DAE Gram matrix of rank~12. Under the assumptions of Corollary~\ref{thm:lifting}, the result is a valid CBC for $\mathcal{C}_{\mathrm{poly}}$ on the considered component of $\mathcal{M}$. Whole-body safety with respect to~\eqref{eq:h_wb} is observed in simulation.\par
\endgroup

\begingroup
\spaceskip=.95\fontdimen2\font plus \fontdimen3\font minus \fontdimen4{RQ1: DAE-aware vs.\ ODE-only CBC (\ref{rq:safety})} The control policy derived from the DAE-aware CBC maintains $h_{\mathrm{wb}}(\x(t)) \ge 0$ throughout with zero violations and no QP infeasibility. As a baseline, we compare against an ODE-only CBC following~\cite{clark2024semialgebraic}, synthesized on~\eqref{eq:ode_manipulator} without verifying the feasibility of~\eqref{eq:dae_cbf_qp}. The ODE-only CBC trajectory violates $h_{\mathrm{wb}} < 0$, i.e., physical link-obstacle penetration (Fig.~\ref{fig:three_link_sim}(a), (b)). Enforcing DAE compatibility prevents the safety violation observed with the ODE-only baseline.\par
\endgroup

\parstart{RQ2: Scalability (\ref{rq:scalability})} Table~\ref{tab:sdp_scaling} compares the per-iteration solve time and problem size. The ODE-to-DAE lifting reduces SDP variables from $18,145$ to $406$. Each iteration of the reduced SOS solves in $41.0$\,s, while direct SOS on the full DAE exceeds the solver's memory limit. The DSOS relaxation makes the full DAE tractable at $655.6$\,s per iteration but remains $16\times$ slower than the reduced ODE. Thus, ODE-to-DAE lifting improves scalability while preserving certificate validity on the considered component of $\mathcal{M}$ under Corollary~\ref{thm:lifting}.

\begin{table}[t]
\caption{Comparison of per-iteration solve time and problem size for SOS and DSOS synthesis on the reduced ODE vs.\ the full DAE.}
\label{tab:sdp_scaling}
\resizebox{\linewidth}{!}{%
\begin{tabular}{lccccc}
\toprule
 & $n_d$ & $\dim(m_s)$ & $Q_s$ vars & Constraints & Time (s) \\
\midrule
Red. (SOS)  & 6  & 28  & 406      & LMI ($28{\times}28$)   & 41.0 \\
Red. (DSOS) & 6  & 28  & 406      & Linear (378)           & 40.2 \\
\midrule
DAE (SOS)   & 18 & 190 & 18{,}145 & LMI ($190{\times}190$) & N/A \\
DAE (DSOS)  & 18 & 190 & 18{,}145 & Linear (17{,}955)      & 655.6 \\

\bottomrule
\end{tabular}}
\end{table}

\section{Conclusion}\vspace{-0.1cm}\label{sec:conclusion}
This paper presented a synthesis framework for DAE control barrier certificates that combines an anchored SOS barrier with Positivstellensatz certificates for compatible-input feasibility. The resulting compatibility condition ensures that safety and algebraic consistency are satisfied simultaneously for both index-1 and higher-index systems. A case study on a 24-dimensional three-link manipulator DAE illustrates safety without link violations. Future work includes improving scalability on higher-dimensional DAEs and extending the framework to neural CBFs.

\bibliographystyle{IEEEtran}
\bibliography{ref}
\end{document}